# Understanding COVID-19 Vaccine Reaction through Comparative Analysis on Twitter

Yuesheng Luo[1] and Mayank Kejriwal[1]

[1] Information Sciences Institute, University of Southern California
kejriwal@isi.edu

**Abstract.** Although multiple COVID-19 vaccines have been available for several months now, vaccine hesitancy continues to be at high levels in the United States. In part, the issue has also become politicized, especially since the presidential election in November. Understanding vaccine hesitancy during this period in the context of social media, including Twitter, can provide valuable guidance both to computational social scientists and policy makers. Rather than studying a single Twitter corpus, this paper takes a novel view of the problem by *comparatively* studying two Twitter datasets collected between two different time periods (one before the election, and the other, a few months after) using the same, carefully controlled data collection and filtering methodology. Our results show that there was a significant shift in discussion from politics to COVID-19 vaccines from fall of 2020 to spring of 2021. By using clustering and machine learning-based methods in conjunction with sampling and qualitative analysis, we uncover several fine-grained reasons for vaccine hesitancy, some of which have become more (or less) important over time. Our results also underscore the intense polarization and politicization of this issue over the last year.

## 1 Background

Although the rapid development and manufacturing of COVID-19 vaccines has been touted as a modern miracle, vaccine hesitancy is still very high in the United States and many other western nations [1, 2]. As of the time of writing (October 2021), even single-dose vaccination in the US has only just crossed 65% and full-dose vaccination is significantly lower[1]. Since vaccine supply currently far outstrips demand across the US, communication and outreach have proven to be valuable tools but thus far, the effects have not been as pronounced as expected. Furthermore, vaccine hesitancy is not uniform across all sociodemographic segments of the US population, suggesting there may be complex causal drivers at play [3]. Therefore, a better understanding of vaccine hesitancy at scale, and the public perception of it, especially on platforms like social media, may be valuable for social scientists and policy experts alike in tackling this problem with effective strategies and incentives.

---

[1] https://usafacts.org/visualizations/covid-vaccine-tracker-states/

Table 1 provides some samples of tweets arguably expressing vaccine hesitancy for a variety of reasons. Already from this sample, we find that there is heterogeneity in the reasoning. For example, some tweets have some religious or even conspiratorial connotations (e.g., 'Mark of the Beast', '666 with nano technology to change our DNA') while others are doing an implicit risk analysis, such as in the very first tweet. Given the volume of tweets on this subject on social media, a natural opportunity arises for combining the right tools with qualitative analysis to understand both the reasons, and the prevalence of those reasons, behind vaccine hesitancy.

**Table 1.** Examples of tweets (from the Fall 2020 dataset collected prior to the US Presidential Election) expressing vaccine hesitancy for a variety of reasons.

| |
|---|
| I'm not getting a vaccine for a virus with a 99.5% survival rate. |
| Cuomo: Americans should be 'very skeptical' about COVID-19 vaccine |
| Gov. Newsom added to concerns about COVID-19 vaccines and said the state will review the safety of any vaccine approved by the Trump administration. |
| "Coronavirus will probably never disappear and a vaccine won't stop it completely, according to Sir Patrick Vallance."Then what on earth are the government waiting for? |
| SOUTH KOREA: Five people have died after getting flu shots in the past week, raising concerns over the vaccine's safety just as the seasonal inoculation program is expanded to head off potential COVID-19 complications. |
| Everyone please retweet. Asymptomatic carriers have NEVER driven the spread of airborne viral disease. From FAUCI himself!!! |
| The covid-19 Vaccine is the Mark of the Beast! The 666 with nano technology to change our DNA! |
| Who else REFUSES to get a covid vaccine? |

In this paper, we propose a methodology for conducting a comparative study on this issue using Twitter data. Specifically, we collect data during a period in October 2020 and also in February 2021 using an identical collection and filtering methodology. We compare and contrast these two corpora in rigorous ways, using both statistical analysis on keywords and hashtags, including keywords that are related to the pharmaceutical organizations developing the vaccines (e.g., Moderna) as well as keywords with emotional connotation. We also aim to understand the causes of vaccine hesitancy, and their fluctuation between these two time periods by using, in addition to these purely statistical and count-based analyses, machine learning methods like clustering and word embeddings, in conjunction with manual labeling of potential vaccine hesitancy reasons that we detect in the tweets.

Analysis of the nature described above has utility, both from a computational social science perspective (understanding COVID-19 vaccine hesitancy as a phenomenon in itself) and from a policy perspective (to counteract vaccine hesitancy with effective communication strategies and incentives). We list below two specific research questions we address in this paper.

### 1.1 Research Questions

- How have units of discussion (such as hashtags and keywords) changed on Twitter from Fall 2020 (prior to the US presidential election) to Spring 2021, especially pertaining to COVID-19 and vaccines, and can these changes be explained by events that have come to the public's attention since?
- Can a combination of machine learning (especially, representation learning and clustering) and qualitative sample-based analysis on tweets help understand some of the finer-grained causes, and prevalence thereof, of vaccine hesitancy?

## 2  Related Work

In the last decade, computational social science has become a mainstream area of study [4-6], especially in using social media and other such datasets to understand complex phenomena at lower cost [7-9]. Twitter studies have been particularly prolific [10-12]. To our knowledge, however, comparative studies of the kind we conduct in this paper have been rarer. One reason is the natural confound of using datasets that have not been collected or processed in a near-identical manner e.g., using the same sets of keywords and preprocessing modules, and so on.

At the same time, vaccine hesitancy has also been an area of research in itself [1, 2], well before COVID-19 [13-15]. COVID-19 vaccines were approved in late 2020 and early 2021 for public use in countries across the world. In May, President Biden announced his goal as getting at least 70% of Americans partially vaccinated against COVID-19 by early July at the latest. Current statistics indicate that this goal has only just been achieved (and then only for a single shot). Low vaccination uptake is not due to supply constraints, but rather, due to vaccine hesitancy among segments of the population. Numerous news articles have documented this phenomena[2], but there have been few rigorous vaccine hesitancy studies (specific to COVID-19) due to the recency of the issue. Our study, which uses grassroots social media data, offers some perspectives on this issue both before and after the election.

The methods in this paper rely on established and time-tested machine learning methods that have been found to work well broadly [16], rather than the very latest methods whose utility has not necessarily been validated outside their evaluation contexts, such as specific benchmarks or applications. Many of the natural language algorithms we use are already pre-packaged in open-source software such as the natural language toolkit (NLTK) or Scikit-Learn [17, 18]. Algorithms such as k-means and word representation learning can be found in many now-classic papers or books on machine learning [16, 19, 20].

---

[2] Two relatively recent examples include https://www.sltrib.com/news/2021/09/27/heres-why-vaccine/  and https://www.nytimes.com/2021/09/24/opinion/vaccines-identity-education.html

## 3 Materials and Methods

### 3.1 Data Collection and Statistics

Using the publicly available Twitter API, we started collecting our first dataset, referred to as *Fall 2020*, on October 19, 2020 and finished collecting the data on October 25, 2020. A total of 371,940 tweets was collected over this period using a set of COVID-19 specific keywords and phrases as input to the API[3]. However, despite this focused collection, not all of these tweets were useful for our study. We filtered the tweets by dropping N/A tweets (which could not be retrieved by the API despite having a tweet ID), removing duplicates[4], and removing non-English tweets. After this filtering, we were left with 128,472 tweets. We pre-processed these tweets by removing URLs, special characters, punctuation and emojis. Hashtags were retained as they are important for our subsequent analysis.

**Table 2.** Dataset statistics (for full corpus, retweets and non-retweets) for the two datasets (Fall 2020 / Spring 2021) described in the main text.

| Data | Number of tweets | Number of unique user IDs | Hashtags (total) | Hashtags (unique) |
|---|---|---|---|---|
| Retweets | 22,224 / 76,883 | 18,201 / 57,450 | 5,953 / 18,057 | 2,076 / 5,527 |
| Non-retweets | 106,248 / 281,128 | 72,304 / 173,902 | 21,062 / 117,590 | 5,132 / 21,600 |
| **Total** | 128,472 / 358,011 | 90,505 / 218,528 | 27,015 / 135,647 | 5,672 / 22,966 |

Our second dataset, referred to as *Spring 2021* was collected using a near-identical methodology (and in particular, the same set of keywords) to facilitate accurate comparisons. A total of 970,708 tweets was collected over the period of collection from February 18, 2021 to February 24, 2021; after pre-processing, 358,011 tweets were retained. Statistics on both datasets are tabulated in Table 2, wherein we show not just the total numbers of tweets in both datasets, but the proportion of tweets that are retweets[5], as well as the number of unique user IDs, the total number of hashtags and the unique number of hashtags.

---

[3] "covid vaccine", "coronavirus vaccine", "china virus vaccine", "covid injection", "covid shot", "kungflu vaccine", "wuhan virus vaccine", "pfizer covid".
[4] These are tweets with the same tweet text. It happens when different users retweet the same text without editing the original tweet text.

Note that, despite using the same collection methodology, the volume of tweets (both before and after pre-processing) in the Spring dataset is almost three times higher than that of the Fall dataset, attesting to the increased relative importance that COVID-19 related topics had assumed in Twitter since our initial data collection and following the election. In keeping with our research questions, we study these changes more systematically in *Results*.

### 3.2 Clustering and Labeling

One of the goals of the paper is to discover and understand the reasons for COVID-19 vaccine hesitancy specifically through the analysis of Twitter data. Since our datasets have hundreds of thousands of tweets, manually reading and categorizing tweets is clearly infeasible. However, fully automated approaches are unlikely to be fruitful either, mainly because Twitter data tends to be noisy.

Instead, we rely on a combination of manual and machine learning approaches. Our first step in discovering structure in the data is to cluster the tweets. However, in order to cluster the tweets, we need to first convert them into vectors. In recent times, *word embedding* techniques have emerged as a particularly powerful class of techniques for converting sequences of text into dense, continuous and low-dimensional vectors. Hence, we rely on these word embeddings for tweet vectorization. Before describing the approach, however, we note that, in order to have meaningful tweet vectors, tweets with fewer than 50 characters are filtered out. After this additional filtering step, 29,951 and 43,879 tweets were found to have been excluded from the Fall 2020 and Spring 2021 datasets in Table 2, respectively.

For each tweet, the text is first tokenized using NLTK's *tokenize* package[6] [17], following which we embed each tweet by using the *fastText* 'bag-of-tricks' model implemented in the *gensim*[7] package [21, 22]. Although fastText provides a pre-trained model based on Wikipedia, we decided to train our own model both due to the fact that social media is noisy and non-grammatical, and also because we wanted to control the dimensions of the word embedding vectors. Specifically, we trained a separate word embedding model for each of the two datasets (Fall 2020 and Spring 2021), setting the vector dimensionality to 25 in each case. Once the word embedding model had been trained, we calculated the tweet embedding by averaging the word embeddings of tokens in the tweets, as is standard practice.

Following the procedure above, a single embedding is obtained per tweet. Using these embeddings, we used the K-Means[8] algorithm [19] to cluster the embeddings in the Fall dataset into 7 clusters, following early explorations. Next, for each cluster, we randomly sampled 100 tweets and manually labeled them with a vaccine hesitancy cause (e.g., 'COVID-19 is common flu') to obtain a small, but reliable, ground-truth for better understanding vaccine hesitancy causes qualitatively. After labeling, we were able to determine six broad, but distinct, reasons that seemed to be significantly impacting vaccine hesitancy expressed on Twitter.

---

[5] Note that a retweet can contain more than just the 'original' tweet (that is being retweeted) since the retweeting user can also add to it.
[6] https://www.nltk.org/api/nltk.tokenize.html
[7] https://radimrehurek.com/gensim/models/fasttext.html
[8] We used the implementation provided in the Python sklearn library: https://scikit-learn.org/stable/modules/generated/sklearn.cluster.KMeans.html. We used the default parameters, with k set to 7, and selected initial centers.

For the Spring dataset, the procedure was very similar, except that we picked k=5 rather than k=7. One reason for doing so is that, in our initial exploration, we sampled 1,000 tweets from the Spring 2021 dataset, and found that only 5 out of 6 causes that we had identified in the Fall 2020 labeling exercise were to be found in that sample. After obtaining the 5 clusters, we again sampled 100 tweets from each cluster, and manually labeled them to counter any kind of structural bias. In fact, in this last step, we uncovered a new sixth reason that had not been present before in the Fall 2020 dataset. We comment both on the excluded classes, as well as the specific reasons we identified through the clustering and labeling process, in *Results*.

## 4   Results

Our first set of results concerns a comparison of the top hashtags and keywords in the Spring and Fall datasets. To compute a ranked list of top hashtags and keywords, we collected the hashtags by searching for the hashtag symbol (#). Next, keywords were collected by tokenizing the whole document using a publicly available term frequency-inverse document frequency (tf-idf) vectorization facility[9] [23]. Common English stop words and task specific stop words were removed. Keywords with similar meanings are grouped together. For example, 'death', 'dies' and 'died' are group into one. More specific grouping rules are described in a table that we provide in an appendix for replicability. The popularity of each hashtag or keyword was computed by calculating the frequency of tweets in which it appeared.

Fig. 1 illustrates ranked list of top 30 hashtags for each dataset, with the 'lines' showing which hashtags in the top 30 terms of the Fall dataset are also in the Spring dataset. We find that there was considerable flux, attesting to the rapid and real-time pace of discussion on Twitter, especially on political matters such as the elections that were ongoing in the state of Bihar in India in the fall (and that were well over by the spring) to the US presidential election debate. We also find that 'trump' is not present in the Spring dataset, likely due to the former president being removed from Twitter in January 2021.

In focusing on the vaccine related tweets, we find that 'lockdown' became more frequent in spring, while, as expected, 'covid19' and 'covidvaccine' continued to be the top hashtags. The hashtag 'astrazeneca' went down due to concerns about the AstraZeneca vaccine that had just emerged. In contrast, 'pfizer', which is not even present in the Fall top-30 list, emerges as a top-5 hashtag in the Spring top-30 list. We also find more vaccine-related hashtags in the top-10 in the Spring dataset due to the success of vaccines in early 2021. The 'moderna' hashtag has also entered the top-30 though it is still trending well below 'pfizer'.

---

[9] Specifically, we used the scikit-learn package in Python: https://scikit-learn.org/stable/modules/generated/sklearn.feature_extraction.text.TfidfVectorizer.html with max_features=2000, and ngram_range from 1 to 2 (i.e. unigrams and bigrams).

Fig. 2, which illustrates a similar plot, but for keywords rather than hashtags, shows similar trends. The keyword 'pfizer' and vaccine-related keywords are even more prominent in Fig. 2 rather than Fig. 1, suggesting that, at least on Twitter, keywords may be better at capturing COVID-19 and vaccine-related matters more than just raw hashtags. We find indeed that even 'johnson' and 'johnsonjohnson' appear as prominent keywords in the Spring top-30 list, although these are not present yet as hashtags in the Fig 1 top-30 list. At the time of writing, the Moderna, Pfizer and Johnson & Johnson vaccines have proved to be the most popular options in Western countries.

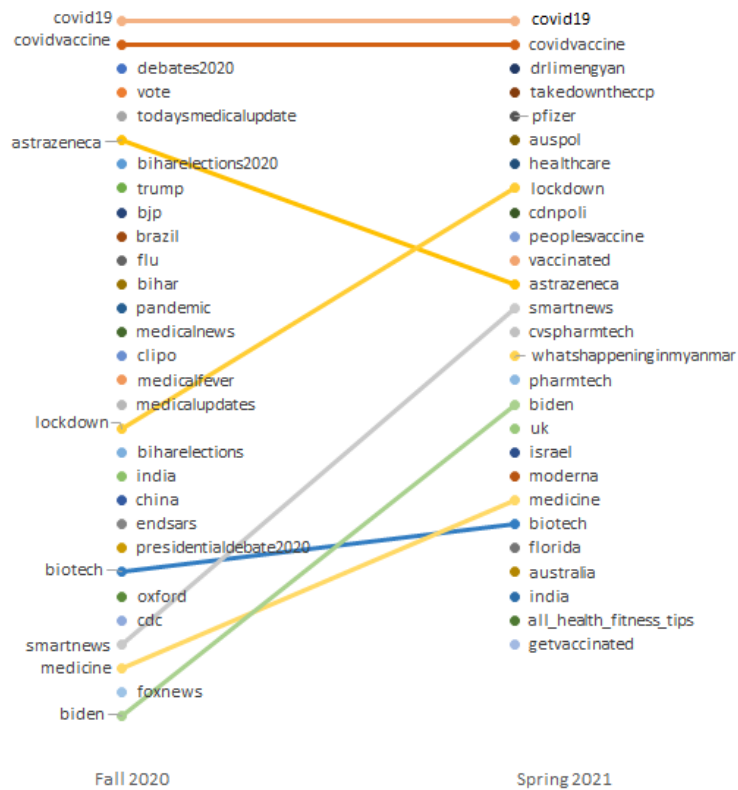

**Fig. 1.** Change in ranked list of top hashtags (in descending order) between the two datasets. The left and right hand columns represent the top hashtags in the Fall 2020 and Spring 2021 datasets respectively.

Fig. 3, 4 and 5 show a slightly different and more quantitative view reinforcing the findings. In Figure 3, we quantitatively map the percentage change of tweets in which a shared keyword (between the Fall 2020 and Spring 2021 datasets) appears. While keywords like 'flu', 'trump' and 'astrazeneca' are very high in the Fall dataset compared to the Spring, the 'pfizer', 'dose', 'vaccinated' and 'vaccination' keywords have become much more prominent in the spring, which also suggests that there is much more discussion about vaccination on Twitter during this time, following the politics-heavy discussions in the fall.

In order to approximately understand the 'public mood' during this time, we also conducted a similar exercise, but with keywords that have an 'emotional connotation'. The high prevalence of words like 'trust', 'lie' and 'concern', which does not *necessarily*[10] indicate either positive or negative sentiment *per se*, suggest intense discussion on vaccines and COVID-19. Unfortunately, we also see a rise in words like 'worry' and 'forced', and a decline in 'hope', but we also witness a sharp rise in 'amazing' and less extremity of prevalence in various words compared to the fall.

---

[10] For example, 'How can I trust the government on vaccines?' and 'I trust the vaccine' both count toward the prevalence of 'trust' but the former indicates more hesitancy and negative sentiment. In *Discussion*, as well as later in this section, we explore sentiments and vaccine-related clusters in more qualitative detail.

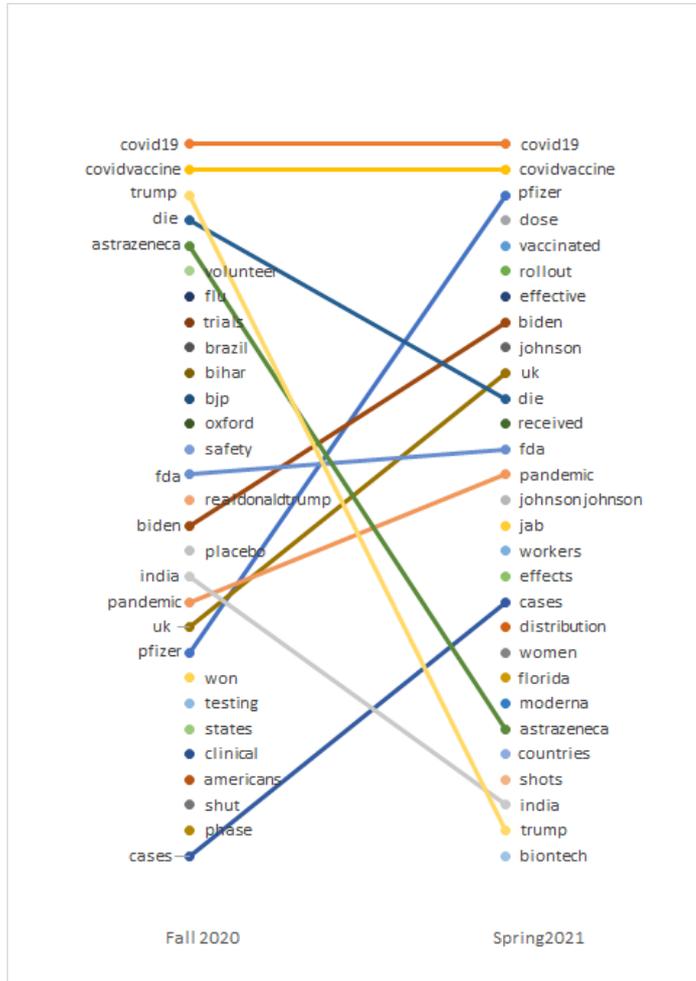

**Fig. 2.** Change in ranked list of top keywords (in descending order) between the two datasets. The left and right hand columns represent the top keywords in the Fall 2020 and Spring 2021 datasets respectively.

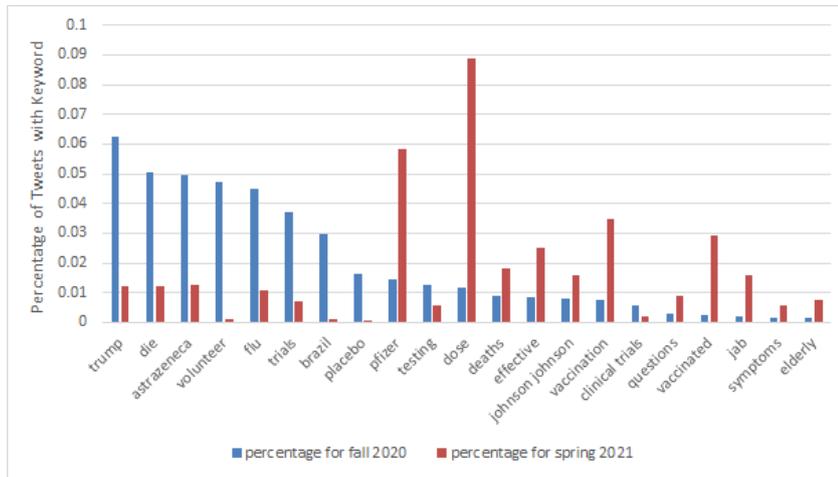

**Fig. 3.** Visualizing the percentage change of tweets in which a shared keyword (between the Fall 2020 and Spring 2021 datasets) appears. Keywords are in descending order of prevalence in the Fall 2020 dataset. We do not consider keywords that are prevalent in at least 0.01% tweets in at least one dataset.

Finally, Fig. 5 compares keywords corresponding to the four vaccine-related organizations on Twitter using the two datasets. As expected, 'astrazeneca' and 'pfizer' witness opposing trends, while 'johnsonjohnson' and 'moderna' remain steady, and with the former exhibiting some growth.

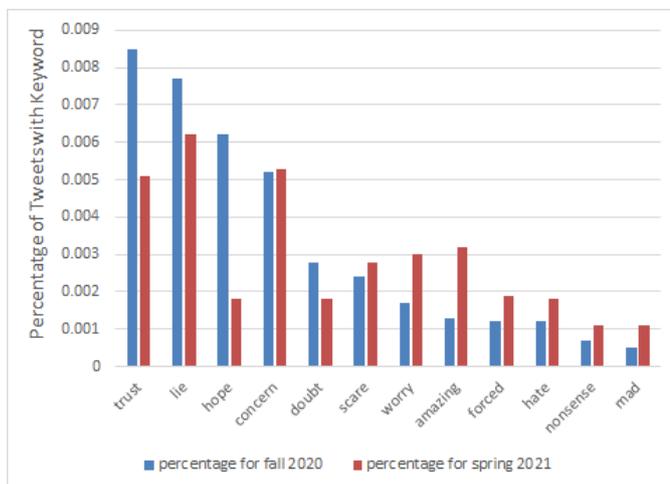

**Fig. 4.** Visualizing the percentage change of tweets in which a shared keyword (between the Fall 2020 and Spring 2021 datasets) with *emotional connotation* appears. Keywords are in descending order of prevalence in the Fall 2020 dataset. We do not consider keywords that are prevalent in at least 0.001% tweets in at least one dataset.

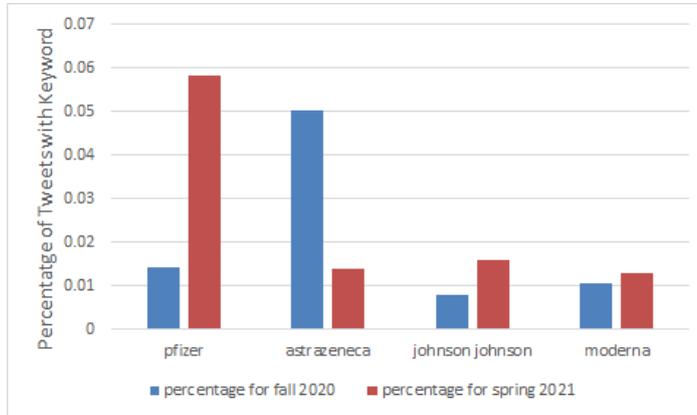

**Fig. 5.** A comparison of the keywords corresponding to the four vaccine-related organizations on Twitter using the two datasets.

In interpreting these results, it must be borne in mind that, although the signals are weaker than might have been expected through direct surveys such as the one conducted by Gallup, the analysis has been conducted over tens of thousands of tweets and users, making it a much larger scale and organic enterprise than a focused and closed-answer survey. Hence, these results should be thought of as being complementary to more rigorous (but smaller-scale and less exploratory) methods, rather than a substitute, in studying COVID-19 from a computational social science lens. Although some of the keywords and hashtags (including the ones with emotional connotation such as in Fig.4) may not always be referring to, or found in, COVID-19 related tweets, our data collection and processing was tailored to such topics to the extent possible. Evidence that many of the tweets are indeed COVID-19 related can be seen both in the prevalence and popularity of COVID-19 related keywords and hashtags earlier in Fig. 1 and 2, as well as in the sample shown in Table 1 and the more qualitative insights discussed subsequently.

**Table 3:** Results from the sampling and labeling exercise for the Fall 2020 / Spring 2021 datasets, the methodology of which was described in *Materials and Methods*.

| Vaccine Hesitancy Reason/Label | Number of positively labeled samples | Percentage of positively labeled samples |
|---|---|---|
| Negative influence | 46 / 0 | 11.1% / 0 |
| Efficacy of the vaccines | 76 / 5 | 18.3% / 10.4% |
| Negative vaccine (trial) news | 174 / 6 | 42% / 12.5% |
| Distrust toward government | 82 / 7 | 19.8% / 14.5% |
| and vaccine research | | |
| Blatantly refuse | 14 / 8 | 3.3% / 16.7% |
| Covid-19 is common flu | 18 / 2 | 4.3% / 4.2% |
| Complaints about vaccine distribution and appointment | 0 / 20 | 0 / 41.7% |
| Total | 414 (out of 700) / 48 (out of 500) | 100% / 100% |

### 4.2 Profiling Vaccine-Related Clusters on Twitter

In *Materials and Methods,* we had proposed a methodology for clustering and labeling the tweets, with results provided below in Table 3. Based on our labeling, the results show that the primary causes of vaccine hesitancy in the fall were negative vaccine news (perhaps caused by the complications observed for the AstraZeneca vaccine), distrust toward government and vaccine research, and vaccine efficacy. After the Biden administration was sworn in, however, we find significant declines across the three causes, perhaps due to better communication, more trust in the administration, and a control of misinformation on social media platforms. However, an unfortunate rise was also observed in the 'blatantly refuse' cluster. Examples of sampled tweets from all the clusters, for both Fall and Spring datasets, are provided in Table 4. Samples for the 'blatantly refuse' cluster suggest a variety of reasons, including conspiracy theories. As more misinformation control measures are implemented, we may see a decline in such clusters; however, this does not necessarily mean that the people who are blatantly refusing have changed their views. Rather, with such control measures, we may have to be more cautious in computing such statistics, as the statistics may be less representative of the population. Since such strong misinformation-control measures have not been implemented on these platforms prior to 2021, we leave an investigation of such selection and sampling-related biases to future work. Some such bias is already observed, since 'negative influence' tweets have virtually disappeared in the Spring dataset despite constituting almost 11% of the Fall dataset. In contrast, the 0% prevalence of tweets in the 'complaints about vaccine distribution and appointment' cluster in the Fall dataset is expected, since vaccines were not available to anyone yet when the Fall dataset was collected.

**Table 4:** Three examples each from the Fall 2020 and Spring 2021 datasets for each of the identified vaccine hesitancy reasons / labels (see Table 3). For the Spring 2021 dataset, note that we could not find any tweets with label *negative influence* likely due to a more concerted effort by social media platforms such as Twitter to crack down on COVID-19 misinformation; similarly, for the Fall 2020 dataset, there were no tweets corresponding to *vaccine distribution & appointment complaints*, probably because viable vaccines had not emerged yet.

| Vaccine Hesitancy | Fall 2020 Dataset | Spring 2021 Dataset |
|---|---|---|

| Reason/Label | | |
|---|---|---|
| Negative Influence | --Cuomo: Americans should be 'very skeptical' about COVID-19 vaccine.<br>--Gov. Newsom added to concerns about COVID-19 vaccines and said the state will review the safety of any vaccine approved by the Trump administration.<br>-- California Governor Newsom is going to have his own health experts review the Covid-19 vaccine which will not make it available until around April. | N/A |
| Efficacy of the vaccines | --"Coronavirus will probably never disappear and a vaccine won't stop it completely, according to Sir Patrick Vallance."Then what on earth are the government waiting for?<br>-- "It is worth reflecting that there's only one human disease that's been truly eradicated"<br>-- If #COVID1984 infection doesn't provide long term immunity then a COVID Vaccine won't either. While it might kill. So our only best hope at moment is population immunity aka herd immunity! | -- The current vaccines won t work against the new covid variant shit vaccine will only speed up covid mutation and make it more deadly<br>-- Local doctor shares concern about covid 19 vaccine side effects for breast cancer survivors<br>-- Avoid the flu shot it dramatically increases your risk of covid if concerned about heart disease change your diet lower your risk to zero most medical orgs and cardiologists are the last you want to listen to or follow unless you want to throw away your health |
| Negative vaccine (trial) news | -- SOUTH KOREA: Five people have died after getting flu shots in the past week, raising concerns over the vaccine's safety just as the seasonal inoculation program is expanded to head off potential COVID-19 complications.<br>-- Brazilian volunteer of AstraZeneca's COVID-19 vaccine trial dies.<br>-- Breaking News: A U.S. government-sponsored clinical trial for a Covid-19 antibody treatment was paused because of a potential safety concern, a day after a vaccine trial was paused after a volunteer fell ill. | -- 4500 people diagnosed with covid after getting 1st vaccine dose<br>-- A 78 year old woman with a history of heart problems died after receiving a covid-19 vaccine in Los Angeles county<br>-- Pfizer vaccine kill woman 78 who died hours after having it via', 'louisiana woman convulsing after pfizer experimental covid vaccine |
| Distrust toward government and vaccine research | -- This article is long but O M G...but PLEASE read it. It gives you all the details of why you MUST resist and learn about WHAT is coming through that COVID19 vaccine. Please, for the future of humanity and your great-grand children, learn and refuse.<br>-- Nope! There is already too much harm done to children, leave them out of this political game!<br>-- IF WE DONT BUY THEIR VACCINE OUR PEOPLE WILL FUCKING DIE | -- Censored Dr Kaufman want to genetically modify us with covid 19 loses his job and willing to go to jail to resist<br>-- Covid live updates minister for regional health urges people in the safest places to still get vaccinated why would he say is he been paid by big pharma to advertise the vaccine to healthy stall cure medicines to symptomatic<br>-- When the globalist make up a pandemic they can lie about the figures up or down now they're saying they're going down only to promote the vaccine |
| Blatantly refuse | -- The covid-19 Vaccine is the Mark of the Beast! The 666 with nano technology to change our DNA!<br>-- no matter bullshit happens never take vid vaccine write fake papers submit bullshit<br>-- Never ever ever ever will I take this vaccine. This should be shared FAR and WIDE | -- Thousands of service members saying no to covid vaccine<br>-- Doctors and nurses giving the covid 19 vaccine will be tried as war criminals<br>-- I wonder about vaccine covid 19 huge toxic hope not force me and parents didn't get this stuff we hate it |
| Covid-19 is common flu | -- I ask this question at least three times a week for the last three months. Why does the world need a vaccine for COVID-19 a.k.a. Chinese Virus that has a recovery rate of 99.74%? No comments huh Bill Gates?<br>-- The science says that children have 99.97% of living with COVID.<br>-- Enough of the Chinese Wuhan virus BS!!! NO way am I taking a vaccine where 99% of the population gets it and, ummm LIVES! | -- Stop making kids wear a mask it's more dangerous for their health than not wearing it their immune systems don't need a vaccine to fight covid-19, covid-21 or any other virus<br>-- boooooo to government of Victoria for locking people down over a fake virus created by china we will not take this fake vaccine to depopulate the population |
| Complaints about vaccine distribution and appointment | N/A | -- The way we are handling the distribution of the covid vaccine is an absolute joke<br>-- The disabled are being systematically denied the covid vaccine across the |

|  |  | country<br>-- The county just cancelled my mom's first covid shot back to square one for a shot so angry 80 and get a shot this is ridiculous makes me want to cry |
| --- | --- | --- |

# 5 Discussion

We supplement the findings in *Results* by using sentiment analysis, along with tweet-sampling, to get a sense of how sentiment around the three public figures during COVID-19 (Trump, Biden and Fauci) fluctuated from fall to spring. We use the VADER sentiment analysis tool due to its ease of use and interpretability. For a given tweet, VADER outputs a score between -1 and +1. Tweets with score greater than 0.05 and less than -0.05 are considered positive and negative, respectively, with the remainder of the tweets labeled as 'neutral'.

Table 5 shows that percentage of positive tweets for all three figures has actually declined and percentage of negative tweets has increased. The numbers reflect the polarization on Twitter, but it is worthwhile noting that the negative percentage of tweets for Trump increased very significantly (by over 20 percentage points) from fall to spring, while positive percentage declined by 5 percentage points. For Fauci and Biden, the numbers are steadier and more consistent. For instance, while positive sentiment for Biden declines significantly, negative sentiment only increases modestly, the opposite trend observed for Trump. More encouragingly, the percentage of tweets that are somewhat neutral remains steady for Fauci, increases for Biden (from 25 to 34) and more than halves for Trump. The results also indicate that, in the Twittersphere, Fauci is as polarizing a figure as the political figures. Accounts in the mainstream media seem to bear this finding out, with Fauci considered overwhelmingly positively and negatively on left-wing and right-wing outlets, respectively. An important agenda is to replicate these analyses using other sentiment analysis tools, as well as comparative analyses using datasets collected after the Spring dataset.

**Table 5:** Sentiment score ratios of tweets mentioning 'Biden', 'Trump' or 'Fauci' using the Vader sentiment analysis tool. We show results for both the Fall 2020 / Spring 2021 datasets.

| Figure | Positive | Negative | Neutral | Total tweets |
| --- | --- | --- | --- | --- |
| Biden | 0.49 / 0.32 | 0.26 / 0.34 | 0.25 / 0.34 | 1,761 / 6,722 |
| Fauci | 0.41 / 0.33 | 0.31 / 0.39 | 0.28 / 0.27 | 679 / 3,165 |
| Trump | 0.40 / 0.35 | 0.31 / 0.54 | 0.29 / 0.11 | 6,088 / 4,297 |

We sample some tweets from both datasets, concerning all three figures, with positive sentiment labels (Table 6) and negative sentiment labels (Table 7). To ensure that we only consider the tweets that are unambiguously positive or negative, we isolate the sets of tweets with VADER sentiment scores greater than 0.9 or less than -0.9, respectively. We sample tweets from these more extreme sets in Tables 6 and 7, respectively. The results show that the tweets are divided along party lines, with other political figures (such as Harris and Cuomo) mentioned as well. Depending on the sentiment, the tweets are divided along the lines of who should take credit for the vaccine, and also whether the vaccine is safe. Negative sentiment tweets tend to have more explicit invectives in them, as expected. While qualitative and small-sample, the tweets highlight the intense polarization around, and politicization of, a public health issue (COVID-19), which is unfortunate.

**Table 6:** Examples of tweets concerning Trump, Fauci and Biden that were labeled as having positive sentiment (score >= 0.9) by the Vader sentiment analysis system. Explicit content has not been filtered, although the tweets have been preprocessed using the methodology presented in *Materials and Methods*.

| **Figure** | Fall 2020 | Spring 2021 |
|---|---|---|
| Biden | -- trump says good chance coronavirus vaccine ready weeks biden predicts dark winter pick president america<br>-- absolute best case biden perfect vaccine rollout gets credited ending coronavirus lol<br>-- let get real people joe biden may greatest guy world blessed thing end covid going another shutdown vaccine ready ready people wear wear masks please nothing change | -- you mean how they take credit for everything that was done by the previous yeah i noticed that too joe biden has done a fabulous job developing the covid 19 vaccine amazing what he has accomplished in a month<br>-- thank you president biden a caring competent president will save america from covid we just wish it could have happened 400 000 folks ago<br>-- oh my god thank you please president biden tells to the lab scientists upgrade some powerful safe effective vaccines |
| Fauci | -- we love care dr fauci voice medically sound knowledge reason would take vaccine vid 19 says safe so<br>-- fauci probably highly respected infectious disease expe world also terrific communicator great confidence francis collins director national institutes health<br>-- fauci predicts safe effective coronavirus vaccine end year | -- fauci says yes to hugs white house chief medical adviser dr anthony fauci says that it s very likely that family members who have been vaccinated against coronavirus can safely hug each other<br>-- love dr fauci so kind so congenial and so committed<br>-- dr fauci has said this my friend hope you have a positive and amazing week |
| Trump | -- wow cant wait coronavirus vaccine stas getting distributed weeks wow god bless president trump<br>-- trump says good chance coronavirus vaccine ready weeks biden predicts dark winter pick president america | -- thanks to the fantastic efforts of president trump for promoting an enviroment creating a vaccine and pushing pharma to be better and also create a path to combat this china virus<br>-- my friend who plagued me with her love of trump for four years is |

|  | -- fda plan ensure safety vaccines developed since trump dismantled office charge trump administration shut vaccine safety office last year | now convinced that the covid vaccine is 666 she was never an anti vaxxer not till now<br>-- south dakota is doing a great job getting out people vaccinated we have a great governor and medical system our medical personnel are doing an average of 900 folks per day and our civil air patrol are getting the vaccine out into smaller towns thanks you president trump |
|---|---|---|

**Table 6:** Examples of tweets concerning Trump, Fauci and Biden that were assigned score less than -0.9 by the Vader sentiment analysis system. Explicit content has not been filtered, although the tweets have been preprocessed using the methodology in *Materials and Methods*.

| Figure | Fall 2020 | Spring 2021 |
|---|---|---|
| Biden | -- cuomo joins biden harris spreading reckless anti vaxxer misinformation right arrow curving down<br>-- biden want close borders scaring people getting vaccine democrats history wanting people die look murder rate cities aboion<br>-- biden keeps lying lying lying lying people eat up | -- joe biden said in his town hall last night that african americans and latinos don t know how to register online to get the vaccine this racist shit along with his lies about children can t spread smdh<br>-- trump says biden is either lying or he s mentally gone after claiming there was no covid vaccine when he took office and floats creating his own social media after twitter ban and teases a 2024 run via fuck you fat basterd |
| Fauci | -- fauci stupid science wrong covid go away vaccine coming election new health care plan wall get built trade deals get better deeper debt world relations worse crisis vote<br>-- fauci fraud needs dealt like traitors criminals need dealt with<br>-- vaccine agenda fail fauci birx cuomo lamont plan fail spent billions terrible idea ignoring treatments work pushing fear | -- and why are trumplicans now mad about vaccines when for 11 months refused to wear masks said covid was no worse than the flu and called people sheep for listening to fauci instead of get to the end of the vaccine line<br>-- anyone who listens to fauci is just stupid the sob lied and people died because of the lies<br>-- dr fauci said he had pain in these 2 places after the covid vaccine dr fauci is a pain all by himself lying to |
| | | the media and scaring people on a regular basis |
| Trump | -- trump fake wife fake economy fake coronavirus vaccine fake insurance plan<br>-- for months trump routinely undermined advice doctors researchers mes fighting covid 19 examines operation warp speed political fight vaccine turmoil inside fda<br>-- trump persists refusing publicly acknowledge true severity coronavirus crisis america he lied failed forewarn public impending danger jan rejecting projections due to | -- imbecile biden ipotus is an idiot he and carmala said trump should have left a stockpile of vaccine what assholes needs to be kept in refrigerated state you f n idiot<br>-- 60 million americans were infected in 2009 with joe biden in charge if you apply that number of infected to covid instead of swine flu in 2009 we would have 1 3 million dead with no vaccine in site be grateful trump developed 6 vaccines with the first from pfizer in 9 months<br>-- covid is a hoax mask do nothing the vaccine are for people not the virus the death rate has not change in 3 year they is no pandemic it is a power and money grab by the globalist and the ccp it has all been done to remove trump from office election fraud is real |

# 6 Conclusion

Media reports, as well as non-partisan observations, have all noted the alarming rise and politicization of public health issues that, on their face, seem politically neutral. At the same time, vaccine hesitancy remains persistent in the population, although the Biden administration has continued to aggressively promote the benefits of vaccination (including making it mandatory for federal employees in the United States). Understanding vaccine reaction and polarization around COVID-19 this last year from a comparative lens, especially given the scale and reach of social media platforms like Twitter, is an important agenda for computational social scientists and policy makers. For the latter, in particular, it may help inform decisions both in the near future as well as from a planning standpoint for future pandemics. Our study took a step in this direction by present both a methodology for conducting such comparative analyses on raw corpora collected from Twitter, and results guided by both quantitative and sample-driven qualitative analyses. In particular, the latter uncovered some reasons for vaccine hesitancy, and found encouragingly that, in the spring, many of the concerns may have been due to logistics-related matters like vaccine distribution and getting appointments. Indeed, in the months since this data was collected, and as vaccine supply exceeded demand, these concerns have been mitigated and vaccination rates have consequently risen.

Nevertheless, almost 15% of sampled tweets in the Spring dataset suggested that vaccine hesitancy was also being caused by mistrust toward the government, among other similarly fundamental reasons. Therefore, significantly more outreach and communication may be necessary in the next few months to reach vaccination rates of 90% or beyond.

# Appendix

**Table A.1:** Manually determined keyword/hashtag groupings used in *Results*.

| | |
|---|---|
| covid19 | covid/covid19/ covid一19/ covid-19/ covid_19/ coronavirus/ covid__19/ corona |
| covidvaccine | covidvaccine/ covidvaccines/coronavirusvaccine/covaxin/ covax/ covid19vaccination/covidvaccination/vaccines/vaccine |
| die | deaths/dies/died |
| dose | dose/doses |
| trust | trust/trusted |
| worry | worried/worries/worrying |
| scare | scary/scared/scare |
| concern | concerns/concerned/concerning |
| lie | lying/lies/lied |
| doubt | doubts/doubt |
| hope | hopes/hopeful |